\documentclass[journal=aamick,manuscript=article]{achemso}

\usepackage[usenames,dvipsnames,table]{xcolor}
\usepackage[version=3]{mhchem} %
\usepackage{miller}
\usepackage{subcaption}
\usepackage{multirow}
\usepackage{color}
\usepackage{textgreek}

\usepackage{amsmath,amssymb}
\graphicspath{{Figures/}}

\usepackage{pdfpages}

\usepackage[%
  breaklinks,             %
  bookmarks = false, %
  pdfpagemode   = UseNone,%
  pdfstartview  = FitH,     %
  pdfstartpage  = 1,      %
  colorlinks    = true,   %
]{hyperref}

\newcommand\myshade{85}
\colorlet{mylinkcolor}{YellowOrange}
\colorlet{myurlcolor}{Aquamarine}
\colorlet{mycitecolor}{violet}

\hypersetup{
  linkcolor  = mylinkcolor!\myshade!black,
  citecolor  = mycitecolor!\myshade!black,
  urlcolor   = myurlcolor!\myshade!black,
  colorlinks = true,
}



\author{Zeeshan Ahmad}
\affiliation{Department of Mechanical Engineering, Texas Tech University, Lubbock, Texas 79409, USA}
\email{zeeahmad@ttu.edu}
\title[xx]
  {A unified and consistent electrical double layer model for treatment of core and space charge layer in solid electrolytes}

\keywords{electrical double layer, defects, space charge, ionic conductivity, interfaces}

\begin{document}

\begin{abstract}
The electrical double layer (EDL) is fundamental to the operation of devices for electrochemical energy storage and conversion. Existing models of EDL in solid electrolytes focus predominantly on the space charge layer and lack a complete treatment of the core layer which is an integral part of the EDL.  The core layer exhibits significant variations in defect properties, such as defect formation energy (DFE), which influence the ionic charge carrier distribution in the space charge layer.   In this work, we develop a general framework for treating both the core and space charge layer in solid electrolytes under dilute and concentrated regimes. We incorporate functional forms of the DFE variation in the core layer and defect-defect interactions that are consistent with previous first-principles simulations of solid electrolytes at surfaces and interfaces. Our simulations reveal that the core layer significantly impacts the potential distribution and defect concentrations in the solid electrolyte. In addition, the core contributes substantially to the conductivity when the surface DFE is much lower than the bulk DFE. Our model paves the way for accurate calculations of properties such as capacitance and ionic conductivity of solid-solid interfaces required for enhancing and optimizing the performance of solid-state electrochemical devices.

\end{abstract}

\section{Introduction}
Electrochemical interfaces involving solid electrolytes play a fundamental role in determining the performance of devices in a variety of applications such as solid-state batteries~\cite{xuInterfacesSolidStateLithium2018}, solid-oxide fuel cells~\cite{boldrinProgressOutlookSolid2019}, electrochemical capacitors~\cite{simonMaterialsElectrochemicalCapacitors2008}, and solar cells~\cite{shaoRoleInterfacesPerovskite2020}. In contrast to solid-liquid interfaces, the structure of the electrical double layer (EDL) and the behavior of charge carriers near solid-solid interfaces remain poorly understood. The standard Guoy-Chapman-Stern theory~\cite{bardElectrochemicalMethodsFundamentals2022} used in liquid electrolytes may not be applicable for solid electrolytes due to large potential drops and distinct capacitive behavior of their interfaces~\cite{swiftModelingElectricalDouble2021,horrocksDiscretenessChargeEffects1999,geDoubleLayerCapacitance2011}. The EDL directly impacts the nature of the potential drop across the interface and the accumulation/depletion of ionic charge carriers in the solid electrolyte. 
Hence, understanding the EDL is essential for optimizing solid electrolytes for enhanced electrical transport and charge transfer in electrochemical devices. 

The EDL is defined as the region near the interface of the solid electrolyte where properties such as ionic carrier concentration and potential deviate from the bulk values. 
This deviation is driven by significant changes in the charge carriers’ properties near the interface, which arise from modified atomic coordination, bonding environments, or local disorder.
This region, referred to as the core or interaction layer exhibits changes in defect properties such as the defect formation energy (DFE)~\cite{jamnikInterfacesSolidIonic1995a}. In addition, defects in the core layer may exist in different configurations and sites such as kinks and surface adions at the interface, leading to different properties compared to the bulk. 
The region between the core and the bulk experiencing the effects of the core layer (e.g. ion accumulation or depletion) is referred to as the space charge layer. 
The core and space charge layers typically exhibit opposing behaviors in terms of charge carrier accumulation and depletion.

While both the core and space charge layers play critical roles in determining the transport properties of solid electrolytes, most studies have focused exclusively on the space charge layer to explain and control these effects~\cite{maierIonicConductionSpace1995,swiftFirstPrinciplesPredictionPotentials2019,mebaneGeneralisedSpacechargeTheory2015,kliewerSpaceChargeIonic1965,braunThermodynamicallyConsistentModel2015,keaneAsymptoticAnalysisSpace2023}. 
Interfaces where charge carriers are accumulated in the space charge layer are believed to lead to enhanced conductivity and vice versa.
Applying this reasoning, low grain boundary conductivity of some solid electrolytes has been thought to be caused by the depletion of mobile lithium (Li) ions in the space charge layer~\cite{wuOriginLowGrain2017,chengRevealingImpactSpaceCharge2020}. A notable departure from this view was observed in the case of $\mathrm{Li_{0.33}La_{0.56}TiO_3}$ where the EDL consists of a Li-deficient core region and a Li-rich space charge layer~\cite{guAtomicscaleStudyClarifying2023}. In this system, the grain boundary conductivity was found to be dominated by the core region, where Li-deficiency leads to reduced conductivity. 
This highlights the importance of considering both the core and space charge layers when modeling the transport behavior of solid electrolytes, as the core layer can significantly influence the overall performance. Moreover, the contribution of the core layer differs depending on whether the transport occurs parallel or perpendicular to the interface~\cite{jamnikInterfacesSolidIonic1995a}.  These insights suggest that discrepancies observed between modeling and experimental properties of the EDL such as impedance, capacitance, and layer thickness may be caused by the absence of explicit treatment of the core layer in current models~\cite{deklerkSpaceChargeLayersAllSolidState2018,geDoubleLayerCapacitance2011,yamamotoDynamicVisualizationElectric2010}.

In this work, we present a unified model of EDL in solid electrolytes with explicit treatment of both the core and space charge layers. We move beyond the existing treatment of the core layer properties as a step function consisting of only a monolayer~\cite{maierIonicConductionSpace1995} and ensure our core layer treatment is consistent with the results of first-principles simulations, notably the variation in DFE of solids near surfaces and interfaces~\cite{limonHeterogeneityPointDefect2024c,ahmadModulationPointDefect2024c}. These simulations revealed that the influence of the core layer can extend across multiple layers, reaching up to 40 {\AA}.
To determine the potential and defect concentration profiles, we employ the local chemical potential-based thermodynamic formulation~\cite{franceschettiLocalThermodynamicFormalism1981}, a powerful tool for modeling equilibrium conditions in solid electrolytes~\cite{franceschettiLocalThermodynamicFormalism1981}.
We find that the DFE variation near the core region significantly impacts the potential drop and the concentration profile of defects. In particular, a large reduction in DFE at the interface leads to higher potential drop across the core region while increase in DFE at the interface results in a lower potential drop across the core region. A similar trend is also found in the capacitance of the interface, leading to higher capacitance when the DFE is lower at the interface. Additionally, we incorporate defect-defect interactions in our model, enabling us to investigate the behavior of solid electrolytes under high defect concentrations. Finally, we estimate the impact of these interactions on the ionic conductivity of the solid electrolyte parallel to the interface. We find that the core layer contributes substantially to the conductivity with low surface DFE and analyzing conductivity purely on the basis of space charge layer may be inaccurate.
Overall, accurate predictions of potential and defect concentration profiles from our EDL model are crucial for determining properties of solid-solid interfaces such as interfacial impedance, ionic transport, and charge transfer rates, which are key to optimizing the performance of solid-state batteries and other electrochemical devices.

\section{Electrical double layer model}

We consider a solid electrolyte with interface at $x=0$ and extending to $x\to \infty$ corresponding to the bulk region. The grand canonical ensemble is employed, where the chemical potential serves as the relevant thermodynamic potential. The electrochemical potential of defects in the electrolyte in the dilute limit is given by~\cite{chenElectrochemomechanicalChargeCarrier2021,maierIonicConductionSpace1995}:
\begin{equation}
    \mu = \mu^0 + kT \ln (c) + z e\phi
\end{equation}
where $\mu_0$ is the standard chemical potential of the defect, $c$ is the defect concentration expressed as a fraction of occupied defect sites, $\phi$ is the potential, $e$ is the elementary charge, $k$ is the Boltzmann constant, $T$ is the temperature, and $z$ is the charge on the defect. For a defect, the standard chemical potential is the DFE~\cite{franceschettiLocalThermodynamicFormalism1981} that can be obtained from first-principles calculations. We modify this electrochemical potential to account for spatial variation of standard chemical potential in the core layer $\mu^0\to \mu^0(x)$, site restriction in configurational entropy expression, and defect-defect interactions, giving
\begin{equation}\label{eq:mu}
    \mu = \underbrace{\mu^0(x)  + ze\phi + \mu^{\text{int}}(c)}_{\text{non-configurational}} + \underbrace{kT \ln \left( \frac{c}{1-c} \right)}_{\text{configurational}}
\end{equation}
where the function $\mu^{\text{int}}(c)$ accounts for changes in electrochemical potential due to defect-defect interactions. Naturally, $\mu^{int}(c)\to 0$ as $c\to 0$ in the dilute limit.

\begin{figure}[htbp]
    \centering
    \includegraphics[width=0.4\textwidth]{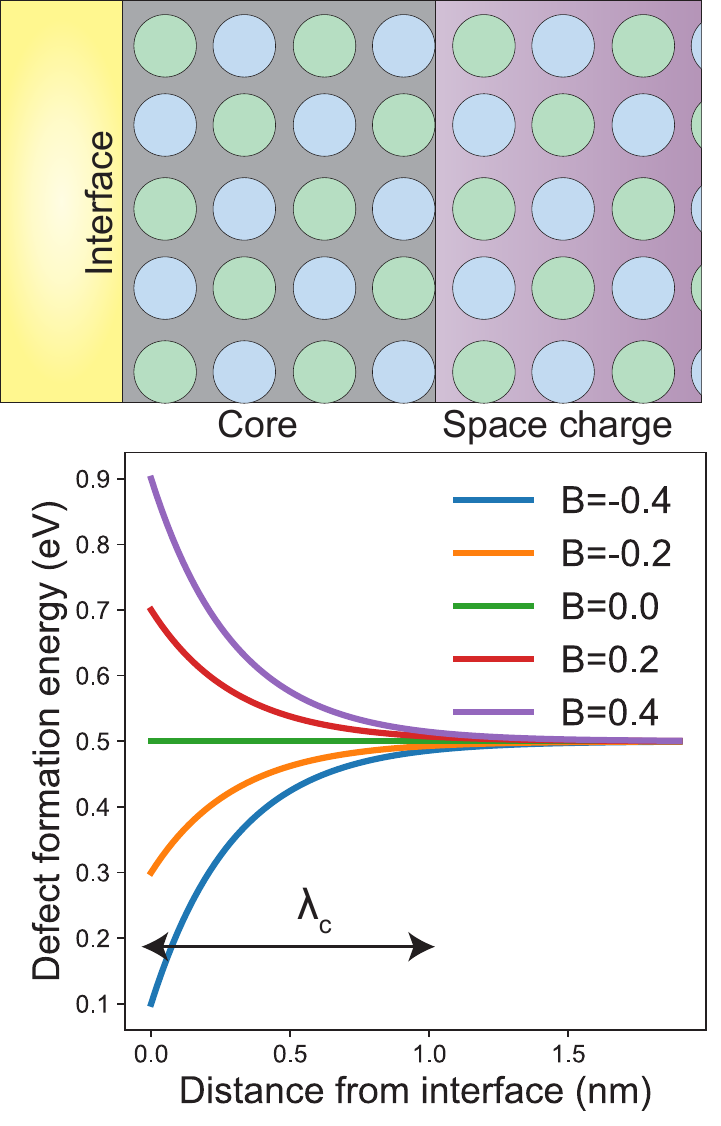}
    \caption{Variation in defect formation energy with distance from the surface/interface. The DFE variation is modeled as a saturating exponential  with a decay length of $\lambda_c$, $DFE = A+B\exp(-x/\lambda_c)$~\cite{limonHeterogeneityPointDefect2024c,ahmadModulationPointDefect2024c}. Here, $A=0.5$ eV and $B$ in the legend is in units of eV. The decay length characterizes the extent of the core layer, which is part of the electrical double layer. $B$ determines the magnitude of DFE variation near the interface compared to the bulk. $B<0$ implies lower surface DFE whereas $B>0$ implies higher surface DFE compared to the bulk.}
    \label{fig:interf}
\end{figure}

The spatially varying standard chemical potential of the defect $\mu^0(x)$ takes the form
\begin{eqnarray}
\mu^0(x) =    \mu^{b0}  + B \exp(-\lambda_c x)
\end{eqnarray}
where $\mu^{b0}$ is the standard chemical potential in the bulk, $B$ is the change in standard chemical potential at the surface compared to the bulk (i.e., the standard chemical potential at the surface is $\mu^{b0} + B$). $\lambda_c$ defines the extent of variation in DFE and can be considered as the length of the core layer.  This form is obtained from our recent first-principles calculations demonstrating an exponential saturating behavior of DFE with distance from the interface for solid electrolytes and semiconductors~\cite{limonHeterogeneityPointDefect2024c,ahmadModulationPointDefect2024c}. \autoref{fig:interf} shows a typical variation in DFE near the interface for bulk DFE of 0.5 eV and different values of $B$. $B<0$ results in lower DFE at the interface leading to enrichment of the interface with defects while $B>0$ results in higher DFE at the interface leading to depletion. Our results for the variation in DFE refine those obtained by \citet{horrocksDiscretenessChargeEffects1999}, who calculated it using an interatomic potential based on Coulomb interactions, in contrast to our first-principles calculations. It is worth noting that this form of DFE variation assumes that the defect type and configuration is similar in the bulk and the interface. It may not be valid when different types of defects are present at the core, e.g., kinks and surface adions. In these cases, first-principles calculations can be used to determine the corresponding DFE variation.

The problem, has two important length scales, $\lambda_c$, the length of the core layer and $\lambda_D$, the Debye length based on the bulk concentrations,  $\lambda_D=\sqrt{\epsilon_r \epsilon_0 k T/ (\Sigma_i N_i z_i^2 e^2 c^b_i)}$. Here, $\epsilon_r $ is the dielectric constant, $\epsilon_0$ is the  permittivity of vacuum, $N_i$ is the site density and $c^b_i$ is the bulk concentration of defect $i$. $\lambda_D$ is often of the order of microns but can be much lower ($<1 $ {\AA}) in case of superionic conductors with high bulk defect concentrations. $\lambda_c$ is usually below 1 {\AA} but could rise to 1 nm depending on the DFE variation in the solid electrolyte. We restrict ourselves to the case where $\lambda_D \gg \lambda_c$ in this work. It would be interesting to explore the EDL in other regimes especially when $\lambda_c\sim \lambda_D$. 

Our electrochemical potential in \autoref{eq:mu} is of the form proposed by \citet{maierIonicElectronicCarriers2001} who divided it into configurational and non-configurational terms. This breakup is useful as the non-configurational term can be treated as an energy level of the ionic charge carriers similar to electronic energy levels. Our formulation contrasts with the one recently proposed by ~\citet{swiftModelingElectricalDouble2021} who used the following expression for defect concentration
\begin{eqnarray}\label{eq:cpbf}
    c = \frac{1}{1/\alpha + \exp[- (\mu - \mu^0 - ze\phi)/kT] }
\end{eqnarray}
This  expression ensures that the defect concentration remains below $\alpha$, the saturated defect concentration, due to defect-defect interactions. Rearranging this expression, we obtain the following form of the electrochemical potential,
\begin{eqnarray}
    \mu = \mu^0 - kT \ln \left(\frac{1}{c} - \frac{1}{\alpha} \right) + ze\phi = \underbrace{\mu^0  + kT \ln \left( \frac{1/c - 1}{1/c-1/\alpha} \right) + ze \phi}_{\text{non-configurational}} +  \underbrace{kT\ln \left( \frac{c}{1-c} \right)}_{\text{configurational}}
\end{eqnarray}
This formulation, while allowing physically reasonable defect concentrations ($c\leq \alpha$), does not provide insights into the nature of the interactions between defects. The physical basis of this form of $\mu^{int}$ is unclear and  $\alpha$ is treated as a parameter than significantly affects properties such as the capacitance of the interface.

Here, we take an alternate approach by controlling the defect interaction strengths in contrast to saturated defect concentration parameter $\alpha$ which is an emergent property. Defect interaction strengths are based on the free energy changes and can be obtained from first-principles simulations for different materials. The chemical potential due to interaction can be derived by using the relation $\mu_{int} = \partial f^{int}/\partial c $ where $f^{int}$ is defect-defect interaction energy. Using finite-size scaling calculations of point defects~\cite{castletonManagingSupercellApproximation2006a,freysoldtFirstprinciplesCalculationsPoint2014}, the free energy of interaction between defects, $f^{int}$ has been found to vary as $\sim L^{-1} + L^{-3}$ where $L$ is the distance between the defects. Since $c \sim L^{-1}$, $f^{int} \sim c + c^3$, and hence $\mu$ varies as $c^2$. This expression is valid for low values of $c$. For higher values of $c$, the interactions should rise rapidly and we require that $\mu^{int}\to \infty$ as $c\to 1$ to prevent material breakdown. This can be accomplished using the function:
\begin{eqnarray}\label{eq:int}
    \mu^{int}(c) =  \frac{f c^2}{(1-c)^2}
\end{eqnarray}
which scales as $c^2$ for small values of $c$. We refer to $f$ as the defect interaction strength.

According to the local chemical potential-based thermodynamic formulation~\cite{franceschettiLocalThermodynamicFormalism1981,maierIonicConductionSpace1995}, the electrochemical potential of the defects must be zero at equilibrium, i.e., $\mu = 0$. Although a solid electrolyte may contain more than two types of defects, it is often the case that the properties of most materials are primarily determined by the interplay of just two dominant oppositely charged defects (Brouwer approximation)~\cite{maierIonicConductionSpace1995}. For simplicity, we consider only these two defects, denoting their properties with subscripts $+$ and $-$. Examples of these defects are Li interstitials and vacancies in a Li-conducting solid electrolyte.
A mean field interaction between these oppositely charged defects of the form $f^{int}=f_{c} c_+ c_-$ can be assumed similar to previous work~\cite{mebaneGeneralisedSpacechargeTheory2015}. We ignore this interaction in this work since this term will be active only when both the defect concentrations are high, which is not the case in our simulations. Then, the electrochemical potential of each defect and the condition for equilibrium may be written as:
\begin{subequations}\label{eq:muboth}
\begin{eqnarray}
    \mu_{+}(x) = \mu_{+}^{b0} - B_+ \exp \left(-\frac{x}{\lambda_{c+}} \right)  + \frac{f_+ c_+^2}{(1-c_+)^2} + f_{c}c^-  + z_+ e \phi + kT \ln \left(  \frac{c_+}{1-c_+}  \right) = 0\\
    \mu_-(x) = \mu_-^{b0} - B_- \exp \left(-\frac{x}{\lambda_{c-}} \right)  + \frac{f_- c_-^2}{(1-c_-)^2} + f_{c}c^+ - z_- e \phi + kT \ln \left( \frac{c_-}{1-c_-} \right)=0  
\end{eqnarray}
\end{subequations}
These equations can be solved together with the Poisson equation for the electrical potential $\phi$,
\begin{eqnarray}\label{eq:poisson}
    \nabla^2 \phi = -e\frac{(N_+ z_+ c_+ - N_- z_- c_-)}{\epsilon_r \epsilon_0}
\end{eqnarray}
Here, $N_+$ and $N_-$ are the site densities of the positively and negatively charged defects.

From symmetry, we arrive at the relationship $d\phi / dx|_{x\to \infty} = 0$, which implies electroneutrality in the bulk, $N_+ z_+ c_+ - N_- z_- c_-=0$. Henceforth, we will focus on solid electrolyte with $z_+=z_-=1$ and $N_+=N_-$. In addition, without loss of generality, we can set the zero of the potential in the bulk of the electrolyte. Using this reference for potential, $\mu^0_+$ and $\mu^0_-$ are replaced by the average value, $A=(\mu^{b0}_+ + \mu^{b0}_-)/2$ (see Supporting Information). $A$ is  the DFE of the solid electrolyte at the charge neutrality condition.

Equation \ref{eq:poisson} needs to be solved with the algebraic constraints \ref{eq:muboth}, together forming a set of differential algebraic equations. We solve these equations using the finite-element method in the open-source Multiphysics Object-Oriented Simulation Environment (MOOSE) framework~\cite{Gaston2009moose}. The equations are solved in a 1D domain from $0\leq x \leq L$. The boundary conditions for the potential are $d\phi/dx|_{x=L}=0$ ($L\gg \lambda_D$) and $\phi(x=0)=-\phi_0$. We use $L=20\lambda_D$ and refer to $\phi_0$ as the EDL potential since it is the total potential drop over the EDL. The code and input file for simulations are available on GitHub~\cite{ahmadedl2024}. 

For concreteness, we assumed a solid electrolyte with values of parameters listed in \autoref{tab:param} for the simulations. The values are based on typical Li-ion conducting solid electrolytes~\cite{swiftModelingElectricalDouble2021,limonHeterogeneityPointDefect2024c}. The resulting Debye length based on bulk defect concentration is 0.19 $\mu$m.  The simulation was performed over a 1D domain of length $L=20\lambda_D$. To reduce the number of parameters, we also assume $B_+=B_-=B$, which implies the DFE of positively and negatively charged defects change by the same amount at the interface. $B$ and $f$ are treated as parameters whose effects on properties are explored. However, they can readily be calculated for each electrolyte interface from first-principles simulations.

\begin{table}[htbp]
    \centering
    \begin{tabular}{|c|c|}
    \hline
    Parameter & value\\
    \hline
        $A=(\mu^{b0}_+ +\mu^{b0}_-)/2$ & 0.5 eV \\
        
        $\epsilon_r$ & 10\\
        $N_+$($=N_-$) & $5\times 10^{28}$ /m$^3$ \\
        $\lambda_{c+}(=\lambda_{c-})$ & 0.3 \AA \\
        $f_c$ & 0 eV\\
        \hline
    \end{tabular}
    \caption{Solid electrolyte parameters used in the simulations.}
    \label{tab:param}
\end{table}

\section{Results and discussion}

\begin{figure}[htbp]
    \centering
    \includegraphics[width=\linewidth]{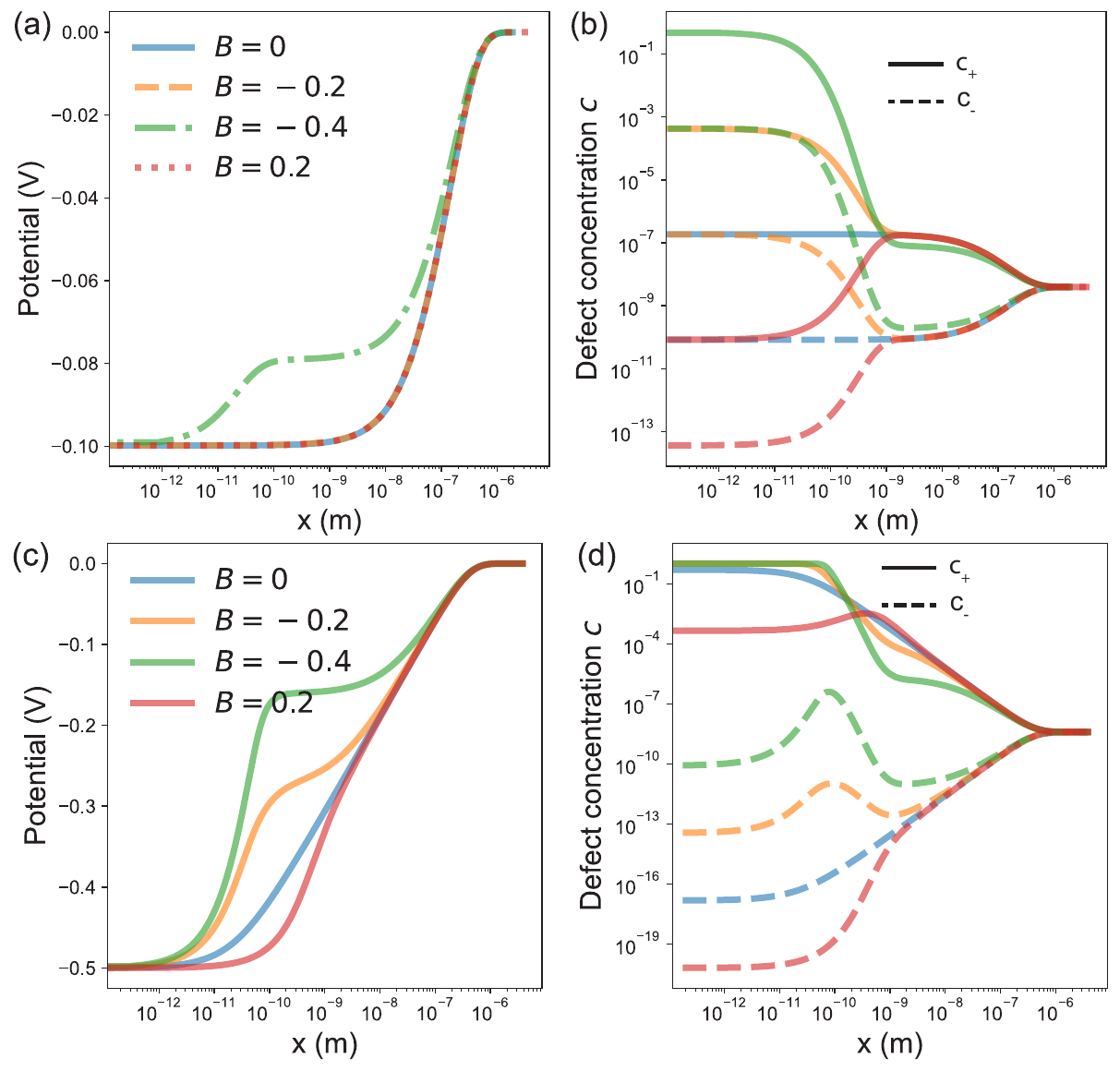}
    \caption{Potential and defect concentration profiles in a solid electrolyte at EDL potentials of 0.1 V (a and b) and 0.5 V (c and d) with and without DFE variation near the interface. The values of $B$ in the legend are in eV. $B=0$ corresponds to no DFE variation, whereas $B\neq 0$ indicates DFE variation near the interface. At low EDL potential of 0.1 V, the potential profiles do not vary substantially with $B$ except at very high magnitude. At high EDL potential of 0.5 V, the potential profile is considerably affected by $B$. The defect concentration profiles deviate significantly from the $B=0$ case near the surface.}
    \label{fig:beffect}
\end{figure}

First, we model the EDL without defect-defect interactions, i.e, $f=0$ eV. The form of the entropy function used $k \log[c/(1-c)]$ still ensures that the defect concentrations do not exceed 1. \autoref{fig:beffect} shows the potential and defect concentration profiles in the solid electrolyte with different variations in DFE near the surface, quantified by the parameter $B$. For a low double layer potential of 0.1 V, the potential profile is affected by the DFE variation only for large negative values of $B$, $-0.4$ eV in this case. Although seemingly large, such values are realistic in solid electrolytes. For instance, in \ce{Li3OCl},  $B$  can be as high as 0.8 eV for Li vacancies.
 At high EDL potential of 0.5 V, $B$ exerts more influence on the potential profile near the interface. More negative $B$ results in a higher potential drop across the core layer ($\sim \lambda_c$) while a positive $B$ decreases the potential drop across the core layer. For example, when $B=-0.4$ eV, more than 60\% of the potential drop occurs across the core layer.

The defect concentration profiles are significantly affected by $B$ near the surface. 
The bulk is electrically neutral with defect concentrations, $c_+=c_- = 3.8 \times 10^{-9}$ in the dilute regime.
For low EDL potential of 0.1 V, we observe an enrichment of the surface with positively charged defects (depletion of negatively charged defects) when $B < 0$, and depletion of positively charged defects (enrichment of negatively charged defects) when $B > 0$, relative to the bulk. At high EDL potential (and at low EDL potential when $B=-0.4$ eV), we observe that the surface defect concentration gets saturated i.e. $c_+(x=0) \sim 1$. This enrichment results in depletion of positively charged defects away from the surface. This observation is consistent with our expectations of opposite behavior in terms of defect enrichment for the core and space charge layers. 
Further, at high EDL potentials, the concentration profile of the negatively charged defect exhibits a maximum for $B<0$ in contrast to monotonic increases when $B=0$. 

Although defect concentrations in the space charge layer and bulk remain in the dilute regime, we observe notably high defect concentrations in the core layer when $B < 0$, which may require the inclusion of defect-defect interactions. This observation contrasts with the work of ~\citet{mebaneGeneralisedSpacechargeTheory2015}, who identified the need for a concentrated EDL theory only  at high dopant concentrations. Our findings emphasize that core layers, even at dilute bulk concentrations, can exhibit behaviors necessitating the treatment of high defect concentrations due to orders of magnitude increase in defect concentration in the core layer. This treatment is essential for correct prediction of properties such as charge transfer rates at the interface which depend sensitively on the interface defect concentration.

\begin{figure}[htbp]
    \centering
    \includegraphics[width=\textwidth]{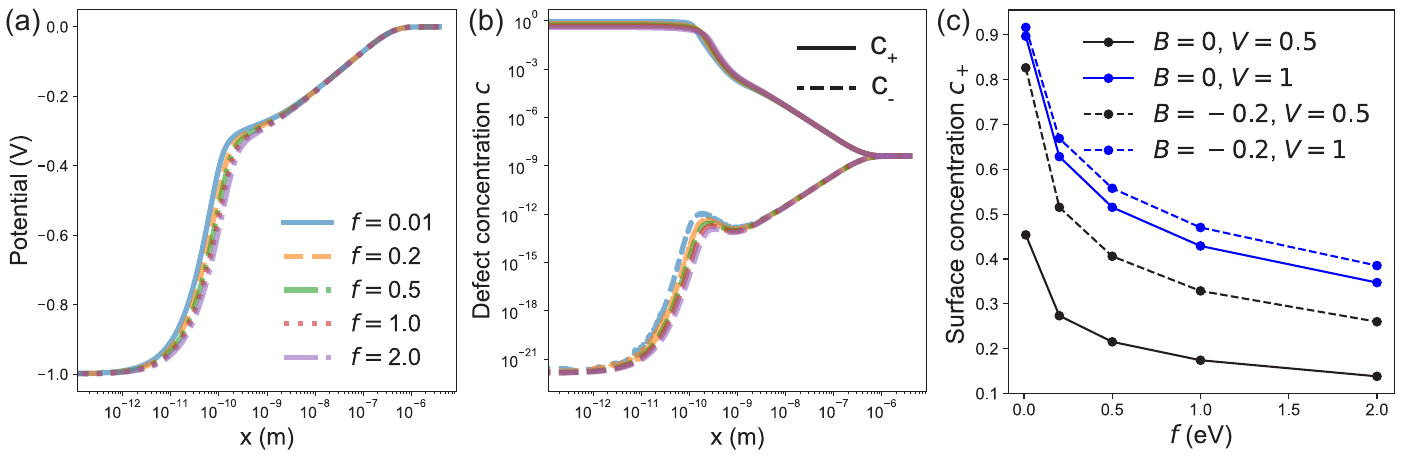}
    \caption{Effect of defect-defect interaction strength, $f$ (in eV) on the potential (a) and defect concentration profiles (b) in a solid electrolyte at double layer potential of 1 V. 
    Both potential and concentration profiles exhibit minor variations. The most significant change occurs in the concentration  of positively charged defect at the surface, $c_+(x=0)$, which decreases as defect-defect interactions increase.}
    \label{fig:int}
\end{figure}

Next, we examine the effect of defect-defect interactions on the potential and defect concentration profiles in solid electrolytes by varying the interaction strength, $f$. According to \autoref{eq:int}, these interactions become significant only at high defect concentrations. 
We find that increase in interaction strengths leads to a lower potential drop across the core layer and higher drop across the space charge layer, as shown in \autoref{fig:int}a. The most pronounced effect of defect-defect interactions is observed in the concentration of positively charged defects at the surface, which decreases monotonically with increasing interaction strength, as seen in \autoref{fig:int}b and c. More negative values of $B$ and higher EDL potentials  result in an increased concentration of positively charged defects at the surface.
Interestingly, the concentration of negatively charged defects, although much lower than the threshold required for defect-defect interactions to get activated, is still influenced by these interactions. This suggests a strong coupling between the concentrations of positively and negatively charged defects, enforced by the Poisson equation.

Using our EDL model, we examine two important properties of solid electrolytes for energy storage applications -  the capacitance versus potential curve of the interface and ionic conductivity parallel to the interface.  The ionic conductivity parallel to the interface is essential to determine the grain boundary conductivity and the conductivity of composite solid electrolytes.

\begin{figure}[htbp]
    \centering
    \includegraphics[width=\textwidth]{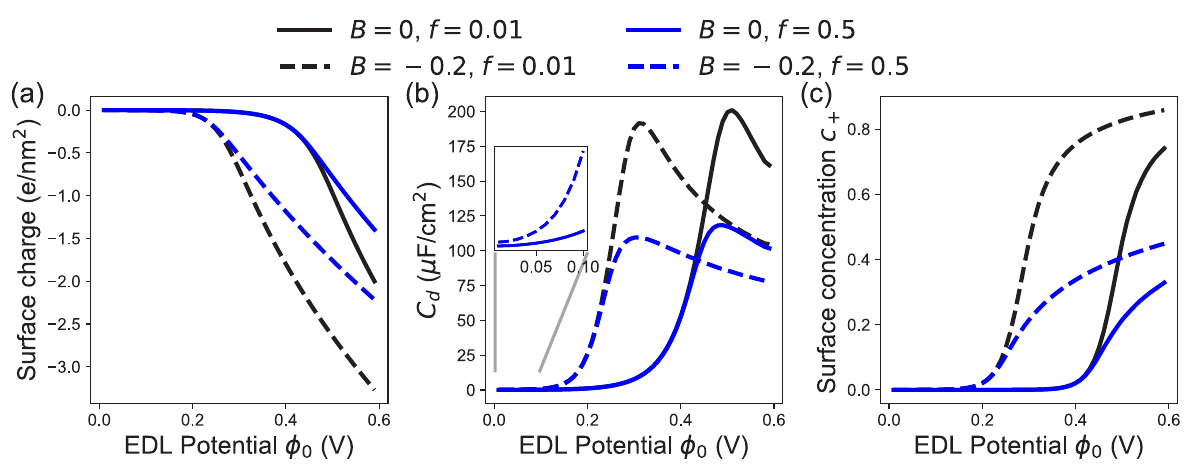}
    \caption{Charge storage properties of the solid electrolyte interface under varying double layer potentials - surface charge, differential capacitance $C_d$, and surface defect concentration of the majority carrier, $c^+$. These properties are plotted at different values of $B$  and $f$ in eV. The surface charge and surface defect concentration are higher in magnitude for $B<0$ due to lower DFE at the surface. The differential capacitance increases faster with double layer potential for $B<0$ with the maxima appearing at a lower potential than the $B=0$ case.}
    \label{fig:capac}
\end{figure}

We calculate the surface charge as $\sigma  = -\epsilon d\phi/dx|_{x=0}$~\cite{bardElectrochemicalMethodsFundamentals2022}. As shown in \autoref{fig:capac}a, the magnitude of surface charge rises slowly  with EDL potential until  a threshold value is reached, beyond which a rapid increase is observed. Furthermore, DFE variation near the interface with $B<0$ results in a higher magnitude of surface charge and a lower threshold EDL potential compared to the uniform DFE case.  The differential capacitance $C_d=d\sigma/d \phi_0$ represents the slope of the $\sigma$ vs. $\phi_0$ plot. At low EDL potentials, the capacitance is nearly zero and then rises sharply to its maximum value. When DFE variation is considered with $B<0$, the capacitance becomes higher than the uniform DFE case near zero EDL potential as seen in the inset of \autoref{fig:capac}b. The inset shows the behavior of capacitance at low EDL potentials and resembles Guoy-Chapman type behavior~\cite{bardElectrochemicalMethodsFundamentals2022} with a higher curvature for $B<0$ compared to $B=0$.
The maximum in capacitance  appears at a lower EDL potential of 0.31 V when $B<0$ compared to 0.51 V for $B=0$. An intriguing result is that the maximum capacitance is higher for $B=0$ than for $B<0$. This may be because the surface concentration of positively charged defects, $c_+(x=0)$ plotted in \autoref{fig:capac}c,  is nearly saturated at its maximum for $B<0$ while it can still increase for $B=0$ as the EDL potential is increased. A key difference in observed in our EDL model compared to ~\citet{swiftFirstPrinciplesPredictionPotentials2019} (\autoref{eq:cpbf}) is that the interface defect concentration continues to rise with $\phi_0$, albeit gradually, as shown in \autoref{fig:capac}c. This contrasts with their results, where the interface defect concentration saturates and reaches a fixed value as the EDL potential is increased. This behavior in our model reflects the influence of defect-defect interactions and the nonlinearity arising from the core layer.

Defect-defect interactions only influence the surface charge and capacitance when  defect concentrations becomes high, which occurs above a certain EDL potential. These interactions reduce the magnitude of surface charge while maintaining the threshold EDL potential for rapid charge increase nearly the same. The capacitance and surface charge reduce considerably due to defect-defect interactions above this EDL potential. However, interactions do not substantially affect the EDL potential at which the maximum in capacitance occurs.

The ionic conductivity of a solid electrolyte is influenced not only by the concentration of defects but also by the migration barrier for ion hopping~\cite{panGeneralMethodPredict2015, linDesignCationTransport2020a}. Similar to the approach used by \citet{maierIonicConductionSpace1995} and \citet{chenElectrochemomechanicalChargeCarrier2021}, we assume that the migration energies are constant and independent of position in order to calculate the ionic conductivity parallel to the interface. This assumption is valid when the EDL potential is small and facilitates a direct comparison with their results. However, this assumption warrants future investigations, as recent work\cite{limonHeterogeneityPointDefect2024c, ahmadInterfacesSolidElectrolyte2021} has demonstrated significant variations in the migration barrier near surfaces and interfaces.

\begin{figure}[htbp]
    \centering
    \includegraphics[width=\textwidth]{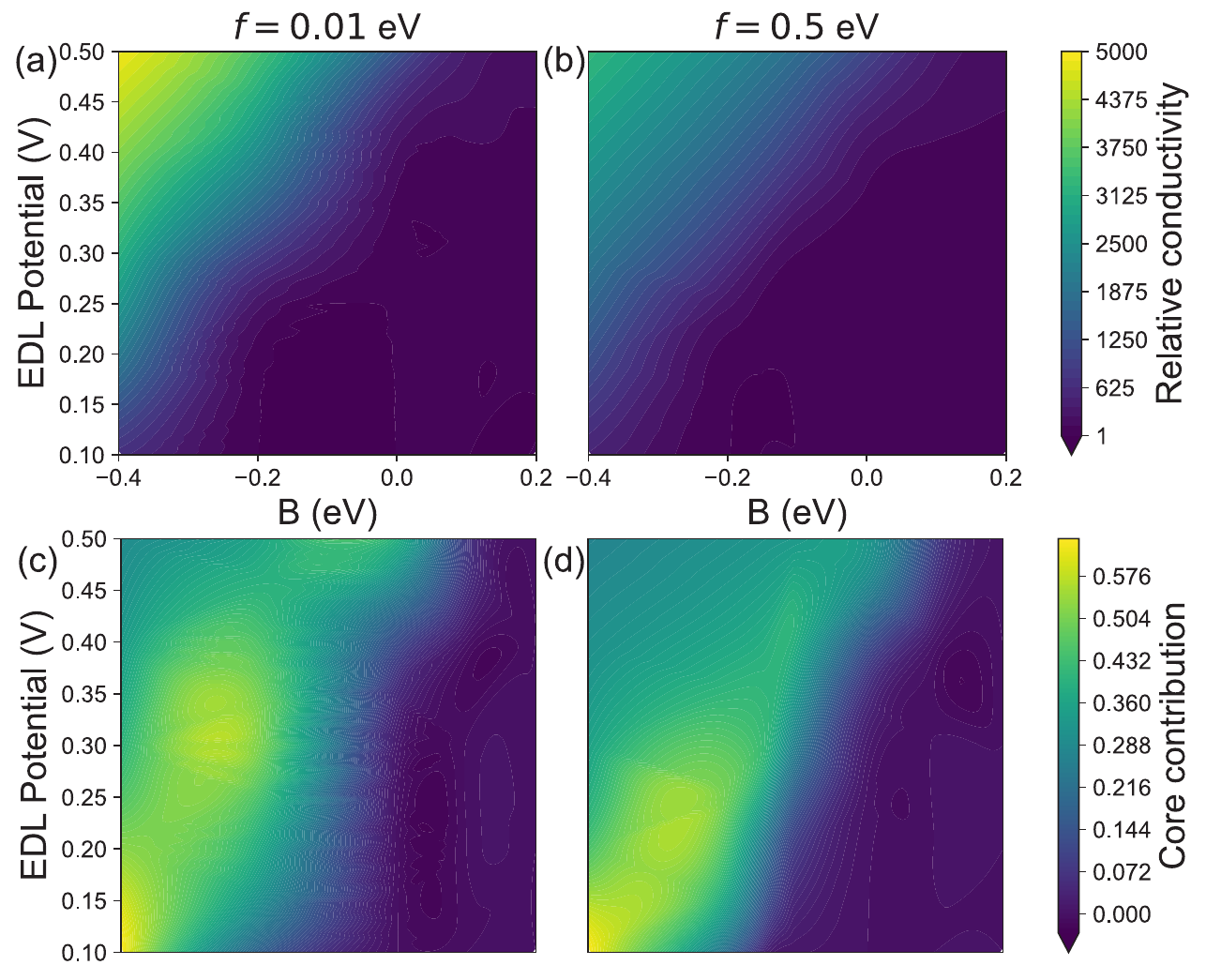}
    \caption{Relative ionic conductivity and core contribution to the conductivity of solid electrolyte as a function of double layer potential and DFE variation quantified through $B$ for two different values of defect-defect interactions $f=0.01$ (a and c) and $0.5$ eV (b and d). Lower $B$ values and higher double-layer potentials result in increased surface defect enrichment, leading to higher conductivity. More negative $B$ values result in higher core contribution to the conductivity.}
    \label{fig:cond}
\end{figure}

Under the constant migration energy assumption, the ionic conductivity is directly proportional to the integral of the defect concentration in the solid electrolyte. This integral can therefore be used to estimate the relative ionic conductivity of the material. We performed the integration over the length of the simulation domain $L=20\lambda_D$. For the current analysis, we assume that ionic transport is carried out by positively charged defects; however, this methodology is applicable to other cases as well. \autoref{fig:cond}a and b map out the  relative conductivity of the solid electrolyte under varying EDL potentials and values of $B$ for interaction strengths, $f=0.01$ eV and $f=0.5$ eV, respectively. The conductivity increases with higher EDL potentials and lower $B$ values, as both factors lead to a higher concentration of positively charged defects near the interface. Additionally, the conductivity is significantly reduced at higher interaction strengths.

Our approach differs fundamentally from previous studies\cite{chenElectrochemomechanicalChargeCarrier2021,mebaneGeneralisedSpacechargeTheory2015}, which solely focus on the conductivity of the space charge layer. In contrast, we explicitly include the  contributions to the conductivity from both the core and the space charge layer. We quantify the contribution of the core layer to the total ionic conductivity using the fraction $r$ of the defects present in the core layer, 
\begin{eqnarray}\label{eq:core}
   r =  \frac{\int_0^{\lambda_c} c_+ dx}{\int_0^L c_+ dx}  
\end{eqnarray}
where $L=20\lambda_D$. \autoref{fig:cond}c and d plot the variation of $r$ for different values of EDL potential and $B$. We find that the contribution of the core layer to the conductivity can be substantial (as high as 50\%) especially when $B<-0.2$ eV regardless of defect interaction strengths. 
We highlight that in cases where the core layer contributes significantly to the conductivity, modifying core layer properties may offer a more effective strategy for enhancing overall conductivity of the solid electrolyte rather than the space charge layer.
It is worth noting that the core conductivity contribution we obtained is a high estimate since discreteness of defect positions in solids~\cite{xiaoDiscreteModelingIonic2022,horrocksDiscretenessChargeEffects1999} and migration energies~\cite{limonHeterogeneityPointDefect2024c} are not included in the analysis. The availability of a  discrete set of positions for defects will reduce the value of the core layer integral in \autoref{eq:core}.
Furthermore, the thickness of the solid electrolyte strongly affects the conductivity contribution of the core. The core contribution decreases with thickness of the solid electrolyte.

Overall, our results emphasize the outsized influence of defects near the interface on the transport properties of solid electrolytes. While the bulk DFE and defect interaction strengths are intrinsic and challenging to manipulate,  the parameters $B$ and $\lambda_c$ can be modified through surface or interface engineering. The values of these parameters can be computed using first-principles simulations and used to screen materials and interfaces to maximize conductivity parallel to the interface. 
This provides a universal and actionable strategy with well-defined design parameters to enhance ionic conductivity of solid electrolytes, enabling tailored performance for diverse solid-state electrochemical applications.

\section{Conclusions}
We have developed a minimal model incorporating the essential physics for simulating the EDL at solid electrolyte interfaces under dilute and concentrated regimes. The model unifies the treatment of core and space charge layers and allows accurate calculation of potential drops, defect concentration profiles, and capacitance of solid electrolyte interfaces. The model effectively incorporates the DFE variation in the core layer near the interface, a key missing ingredient in previous work which can be calculated from first-principles simulations. The core region exerts a strong influence on the conductivity when $B<0$, hence the space charge layer alone may be insufficient for explaining conductivity of grain boundaries and of composite solid electrolytes. 
Overall, our EDL model provides a robust method of predicting ion transport and charge transfer properties of solid electrolytes.

\begin{acknowledgement}
We acknowledge the Texas Tech University Mechanical Engineering startup grant and the High-Performance Computing Center (HPCC) at Texas Tech University for providing computational resources that have contributed to the research results reported within this paper. 
\end{acknowledgement}

\begin{suppinfo}
Derivation of chemical potentials when potential reference is changed

\end{suppinfo}

\bibliography{refs,manual-refs}

\providecommand{\latin}[1]{#1}
\makeatletter
\providecommand{\doi}
  {\begingroup\let\do\@makeother\dospecials
  \catcode`\{=1 \catcode`\}=2 \doi@aux}
\providecommand{\doi@aux}[1]{\endgroup\texttt{#1}}
\makeatother
\providecommand*\mcitethebibliography{\thebibliography}
\csname @ifundefined\endcsname{endmcitethebibliography}
  {\let\endmcitethebibliography\endthebibliography}{}
\begin{mcitethebibliography}{34}
\providecommand*\natexlab[1]{#1}
\providecommand*\mciteSetBstSublistMode[1]{}
\providecommand*\mciteSetBstMaxWidthForm[2]{}
\providecommand*\mciteBstWouldAddEndPuncttrue
  {\def\EndOfBibitem{\unskip.}}
\providecommand*\mciteBstWouldAddEndPunctfalse
  {\let\EndOfBibitem\relax}
\providecommand*\mciteSetBstMidEndSepPunct[3]{}
\providecommand*\mciteSetBstSublistLabelBeginEnd[3]{}
\providecommand*\EndOfBibitem{}
\mciteSetBstSublistMode{f}
\mciteSetBstMaxWidthForm{subitem}{(\alph{mcitesubitemcount})}
\mciteSetBstSublistLabelBeginEnd
  {\mcitemaxwidthsubitemform\space}
  {\relax}
  {\relax}

\bibitem[Xu \latin{et~al.}(2018)Xu, Tang, Cheng, Wang, Liang, Liu, Cao, Wei,
  and Mai]{xuInterfacesSolidStateLithium2018}
Xu,~L.; Tang,~S.; Cheng,~Y.; Wang,~K.; Liang,~J.; Liu,~C.; Cao,~Y.-C.; Wei,~F.;
  Mai,~L. Interfaces in {{Solid-State Lithium Batteries}}. \emph{Joule}
  \textbf{2018}, \emph{2}, 1991--2015\relax
\mciteBstWouldAddEndPuncttrue
\mciteSetBstMidEndSepPunct{\mcitedefaultmidpunct}
{\mcitedefaultendpunct}{\mcitedefaultseppunct}\relax
\EndOfBibitem
\bibitem[Boldrin and Brandon(2019)Boldrin, and
  Brandon]{boldrinProgressOutlookSolid2019}
Boldrin,~P.; Brandon,~N.~P. Progress and Outlook for Solid Oxide Fuel Cells for
  Transportation Applications. \emph{Nature Catalysis} \textbf{2019}, \emph{2},
  571--577\relax
\mciteBstWouldAddEndPuncttrue
\mciteSetBstMidEndSepPunct{\mcitedefaultmidpunct}
{\mcitedefaultendpunct}{\mcitedefaultseppunct}\relax
\EndOfBibitem
\bibitem[Simon and Gogotsi(2008)Simon, and
  Gogotsi]{simonMaterialsElectrochemicalCapacitors2008}
Simon,~P.; Gogotsi,~Y. Materials for Electrochemical Capacitors. \emph{Nature
  Materials} \textbf{2008}, \emph{7}, 845--854\relax
\mciteBstWouldAddEndPuncttrue
\mciteSetBstMidEndSepPunct{\mcitedefaultmidpunct}
{\mcitedefaultendpunct}{\mcitedefaultseppunct}\relax
\EndOfBibitem
\bibitem[Shao and Loi(2020)Shao, and Loi]{shaoRoleInterfacesPerovskite2020}
Shao,~S.; Loi,~M.~A. The {{Role}} of the {{Interfaces}} in {{Perovskite Solar
  Cells}}. \emph{Advanced Materials Interfaces} \textbf{2020}, \emph{7},
  1901469\relax
\mciteBstWouldAddEndPuncttrue
\mciteSetBstMidEndSepPunct{\mcitedefaultmidpunct}
{\mcitedefaultendpunct}{\mcitedefaultseppunct}\relax
\EndOfBibitem
\bibitem[Bard \latin{et~al.}(2022)Bard, Faulkner, and
  White]{bardElectrochemicalMethodsFundamentals2022}
Bard,~A.~J.; Faulkner,~L.~R.; White,~H.~S. \emph{Electrochemical {{Methods}}:
  {{Fundamentals}} and {{Applications}}}; John Wiley \& Sons, 2022\relax
\mciteBstWouldAddEndPuncttrue
\mciteSetBstMidEndSepPunct{\mcitedefaultmidpunct}
{\mcitedefaultendpunct}{\mcitedefaultseppunct}\relax
\EndOfBibitem
\bibitem[Swift \latin{et~al.}(2021)Swift, Swift, and
  Qi]{swiftModelingElectricalDouble2021}
Swift,~M.~W.; Swift,~J.~W.; Qi,~Y. Modeling the Electrical Double Layer at
  Solid-State Electrochemical Interfaces. \emph{Nature Computational Science}
  \textbf{2021}, \emph{1}, 212--220\relax
\mciteBstWouldAddEndPuncttrue
\mciteSetBstMidEndSepPunct{\mcitedefaultmidpunct}
{\mcitedefaultendpunct}{\mcitedefaultseppunct}\relax
\EndOfBibitem
\bibitem[Horrocks and Armstrong(1999)Horrocks, and
  Armstrong]{horrocksDiscretenessChargeEffects1999}
Horrocks,~B.~R.; Armstrong,~R.~D. Discreteness of {{Charge Effects}} on the
  {{Double Layer Structure}} at the {{Metal}}/{{Solid Electrolyte Interface}}.
  \emph{The Journal of Physical Chemistry B} \textbf{1999}, \emph{103},
  11332--11338\relax
\mciteBstWouldAddEndPuncttrue
\mciteSetBstMidEndSepPunct{\mcitedefaultmidpunct}
{\mcitedefaultendpunct}{\mcitedefaultseppunct}\relax
\EndOfBibitem
\bibitem[Ge \latin{et~al.}(2011)Ge, Fu, and Chan]{geDoubleLayerCapacitance2011}
Ge,~X.; Fu,~C.; Chan,~S.~H. Double Layer Capacitance of Anode/Solid-Electrolyte
  Interfaces. \emph{Physical Chemistry Chemical Physics} \textbf{2011},
  \emph{13}, 15134\relax
\mciteBstWouldAddEndPuncttrue
\mciteSetBstMidEndSepPunct{\mcitedefaultmidpunct}
{\mcitedefaultendpunct}{\mcitedefaultseppunct}\relax
\EndOfBibitem
\bibitem[Jamnik \latin{et~al.}(1995)Jamnik, Maier, and
  Pejovnik]{jamnikInterfacesSolidIonic1995a}
Jamnik,~J.; Maier,~J.; Pejovnik,~S. Interfaces in Solid Ionic Conductors:
  {{Equilibrium}} and Small Signal Picture. \emph{Solid State Ionics}
  \textbf{1995}, \emph{75}, 51--58\relax
\mciteBstWouldAddEndPuncttrue
\mciteSetBstMidEndSepPunct{\mcitedefaultmidpunct}
{\mcitedefaultendpunct}{\mcitedefaultseppunct}\relax
\EndOfBibitem
\bibitem[Maier(1995)]{maierIonicConductionSpace1995}
Maier,~J. Ionic Conduction in Space Charge Regions. \emph{Progress in Solid
  State Chemistry} \textbf{1995}, \emph{23}, 171--263\relax
\mciteBstWouldAddEndPuncttrue
\mciteSetBstMidEndSepPunct{\mcitedefaultmidpunct}
{\mcitedefaultendpunct}{\mcitedefaultseppunct}\relax
\EndOfBibitem
\bibitem[Swift and Qi(2019)Swift, and
  Qi]{swiftFirstPrinciplesPredictionPotentials2019}
Swift,~M.~W.; Qi,~Y. First-{{Principles Prediction}} of {{Potentials}} and
  {{Space-Charge Layers}} in {{All-Solid-State Batteries}}. \emph{Physical
  Review Letters} \textbf{2019}, \emph{122}, 167701\relax
\mciteBstWouldAddEndPuncttrue
\mciteSetBstMidEndSepPunct{\mcitedefaultmidpunct}
{\mcitedefaultendpunct}{\mcitedefaultseppunct}\relax
\EndOfBibitem
\bibitem[Mebane and Souza(2015)Mebane, and
  Souza]{mebaneGeneralisedSpacechargeTheory2015}
Mebane,~D.~S.; Souza,~R. A.~D. A Generalised Space-Charge Theory for Extended
  Defects in Oxygen-Ion Conducting Electrolytes: From Dilute to Concentrated
  Solid Solutions. \emph{Energy \& Environmental Science} \textbf{2015},
  \emph{8}, 2935--2940\relax
\mciteBstWouldAddEndPuncttrue
\mciteSetBstMidEndSepPunct{\mcitedefaultmidpunct}
{\mcitedefaultendpunct}{\mcitedefaultseppunct}\relax
\EndOfBibitem
\bibitem[Kliewer and Koehler(1965)Kliewer, and
  Koehler]{kliewerSpaceChargeIonic1965}
Kliewer,~K.~L.; Koehler,~J.~S. Space {{Charge}} in {{Ionic Crystals}}. {{I}}.
  {{General Approach}} with {{Application}} to {{NaCl}}. \emph{Physical Review}
  \textbf{1965}, \emph{140}, A1226--A1240\relax
\mciteBstWouldAddEndPuncttrue
\mciteSetBstMidEndSepPunct{\mcitedefaultmidpunct}
{\mcitedefaultendpunct}{\mcitedefaultseppunct}\relax
\EndOfBibitem
\bibitem[Braun \latin{et~al.}(2015)Braun, Yada, and
  Latz]{braunThermodynamicallyConsistentModel2015}
Braun,~S.; Yada,~C.; Latz,~A. Thermodynamically {{Consistent Model}} for
  {{Space-Charge-Layer Formation}} in a {{Solid Electrolyte}}. \emph{The
  Journal of Physical Chemistry C} \textbf{2015}, \emph{119},
  22281--22288\relax
\mciteBstWouldAddEndPuncttrue
\mciteSetBstMidEndSepPunct{\mcitedefaultmidpunct}
{\mcitedefaultendpunct}{\mcitedefaultseppunct}\relax
\EndOfBibitem
\bibitem[Keane and Moyles(2023)Keane, and
  Moyles]{keaneAsymptoticAnalysisSpace2023}
Keane,~L.~M.; Moyles,~I.~R. An {{Asymptotic Analysis}} of {{Space Charge
  Layers}} in a {{Mathematical Model}} of a {{Solid Electrolyte}}. 2023\relax
\mciteBstWouldAddEndPuncttrue
\mciteSetBstMidEndSepPunct{\mcitedefaultmidpunct}
{\mcitedefaultendpunct}{\mcitedefaultseppunct}\relax
\EndOfBibitem
\bibitem[Wu and Guo(2017)Wu, and Guo]{wuOriginLowGrain2017}
Wu,~J.-F.; Guo,~X. Origin of the Low Grain Boundary Conductivity in Lithium Ion
  Conducting Perovskites: {{Li3xLa0}}.67-{{xTiO3}}. \emph{Physical Chemistry
  Chemical Physics} \textbf{2017}, \emph{19}, 5880--5887\relax
\mciteBstWouldAddEndPuncttrue
\mciteSetBstMidEndSepPunct{\mcitedefaultmidpunct}
{\mcitedefaultendpunct}{\mcitedefaultseppunct}\relax
\EndOfBibitem
\bibitem[Cheng \latin{et~al.}(2020)Cheng, Liu, Ganapathy, Li, Li, Zhang, He,
  Zhou, and Wagemaker]{chengRevealingImpactSpaceCharge2020}
Cheng,~Z.; Liu,~M.; Ganapathy,~S.; Li,~C.; Li,~Z.; Zhang,~X.; He,~P.; Zhou,~H.;
  Wagemaker,~M. Revealing the {{Impact}} of {{Space-Charge Layers}} on the
  {{Li-Ion Transport}} in {{All-Solid-State Batteries}}. \emph{Joule}
  \textbf{2020}, \emph{4}, 1311--1323\relax
\mciteBstWouldAddEndPuncttrue
\mciteSetBstMidEndSepPunct{\mcitedefaultmidpunct}
{\mcitedefaultendpunct}{\mcitedefaultseppunct}\relax
\EndOfBibitem
\bibitem[Gu \latin{et~al.}(2023)Gu, Ma, Zhu, Liu, Wang, Nan, Li, and
  Ma]{guAtomicscaleStudyClarifying2023}
Gu,~Z.; Ma,~J.; Zhu,~F.; Liu,~T.; Wang,~K.; Nan,~C.-W.; Li,~Z.; Ma,~C.
  Atomic-Scale Study Clarifying the Role of Space-Charge Layers in a
  {{Li-ion-conducting}} Solid Electrolyte. \emph{Nature Communications}
  \textbf{2023}, \emph{14}, 1632\relax
\mciteBstWouldAddEndPuncttrue
\mciteSetBstMidEndSepPunct{\mcitedefaultmidpunct}
{\mcitedefaultendpunct}{\mcitedefaultseppunct}\relax
\EndOfBibitem
\bibitem[{de Klerk} and Wagemaker(2018){de Klerk}, and
  Wagemaker]{deklerkSpaceChargeLayersAllSolidState2018}
{de Klerk},~N. J.~J.; Wagemaker,~M. Space-{{Charge Layers}} in
  {{All-Solid-State Batteries}}; {{Important}} or {{Negligible}}? \emph{ACS
  Applied Energy Materials} \textbf{2018}, acsaem.8b01141\relax
\mciteBstWouldAddEndPuncttrue
\mciteSetBstMidEndSepPunct{\mcitedefaultmidpunct}
{\mcitedefaultendpunct}{\mcitedefaultseppunct}\relax
\EndOfBibitem
\bibitem[Yamamoto \latin{et~al.}(2010)Yamamoto, Iriyama, Asaka, Hirayama,
  Fujita, Fisher, Nonaka, Sugita, and
  Ogumi]{yamamotoDynamicVisualizationElectric2010}
Yamamoto,~K.; Iriyama,~Y.; Asaka,~T.; Hirayama,~T.; Fujita,~H.; Fisher,~C.
  A.~J.; Nonaka,~K.; Sugita,~Y.; Ogumi,~Z. Dynamic {{Visualization}} of the
  {{Electric Potential}} in an {{All-Solid-State Rechargeable Lithium
  Battery}}. \emph{Angewandte Chemie International Edition} \textbf{2010},
  \emph{49}, 4414--4417\relax
\mciteBstWouldAddEndPuncttrue
\mciteSetBstMidEndSepPunct{\mcitedefaultmidpunct}
{\mcitedefaultendpunct}{\mcitedefaultseppunct}\relax
\EndOfBibitem
\bibitem[Limon and Ahmad(2024)Limon, and
  Ahmad]{limonHeterogeneityPointDefect2024c}
Limon,~M. S.~R.; Ahmad,~Z. Heterogeneity in {{Point Defect Distribution}} and
  {{Mobility}} in {{Solid Ion Conductors}}. \emph{ACS Applied Materials \&
  Interfaces} \textbf{2024}, \emph{16}, 50948--50960\relax
\mciteBstWouldAddEndPuncttrue
\mciteSetBstMidEndSepPunct{\mcitedefaultmidpunct}
{\mcitedefaultendpunct}{\mcitedefaultseppunct}\relax
\EndOfBibitem
\bibitem[Ahmad \latin{et~al.}(2024)Ahmad, Limon, and
  Ahmad]{ahmadModulationPointDefect2024c}
Ahmad,~B.; Limon,~M. S.~R.; Ahmad,~Z. Modulation of Point Defect Properties
  near Surfaces in Metal Halide Perovskites. \emph{Physical Review Materials}
  \textbf{2024}, \emph{8}, 125402\relax
\mciteBstWouldAddEndPuncttrue
\mciteSetBstMidEndSepPunct{\mcitedefaultmidpunct}
{\mcitedefaultendpunct}{\mcitedefaultseppunct}\relax
\EndOfBibitem
\bibitem[Franceschetti(1981)]{franceschettiLocalThermodynamicFormalism1981}
Franceschetti,~D. Local Thermodynamic Formalism for Space Charge in Ionic
  Crystals. \emph{Solid State Ionics} \textbf{1981}, \emph{2}, 39--42\relax
\mciteBstWouldAddEndPuncttrue
\mciteSetBstMidEndSepPunct{\mcitedefaultmidpunct}
{\mcitedefaultendpunct}{\mcitedefaultseppunct}\relax
\EndOfBibitem
\bibitem[Chen \latin{et~al.}(2021)Chen, Yin, Kang, Cai, and
  Chueh]{chenElectrochemomechanicalChargeCarrier2021}
Chen,~C.-C.; Yin,~Y.; Kang,~S.~D.; Cai,~W.; Chueh,~W.~C.
  Electro-Chemo-Mechanical Charge Carrier Equilibrium at Interfaces.
  \emph{Physical Chemistry Chemical Physics} \textbf{2021}, \emph{23},
  23730--23740\relax
\mciteBstWouldAddEndPuncttrue
\mciteSetBstMidEndSepPunct{\mcitedefaultmidpunct}
{\mcitedefaultendpunct}{\mcitedefaultseppunct}\relax
\EndOfBibitem
\bibitem[Maier(2001)]{maierIonicElectronicCarriers2001}
Maier,~J. Ionic and Electronic Carriers in Solids---Physical and Chemical Views
  of the Equilibrium Situation. \emph{Solid State Ionics} \textbf{2001},
  \emph{143}, 17--23\relax
\mciteBstWouldAddEndPuncttrue
\mciteSetBstMidEndSepPunct{\mcitedefaultmidpunct}
{\mcitedefaultendpunct}{\mcitedefaultseppunct}\relax
\EndOfBibitem
\bibitem[Castleton \latin{et~al.}(2006)Castleton, H{\"o}glund, and
  Mirbt]{castletonManagingSupercellApproximation2006a}
Castleton,~C. W.~M.; H{\"o}glund,~A.; Mirbt,~S. Managing the Supercell
  Approximation for Charged Defects in Semiconductors: {{Finite-size}} Scaling,
  Charge Correction Factors, the Band-Gap Problem, and the Ab Initio Dielectric
  Constant. \emph{Physical Review B} \textbf{2006}, \emph{73}, 035215\relax
\mciteBstWouldAddEndPuncttrue
\mciteSetBstMidEndSepPunct{\mcitedefaultmidpunct}
{\mcitedefaultendpunct}{\mcitedefaultseppunct}\relax
\EndOfBibitem
\bibitem[Freysoldt \latin{et~al.}(2014)Freysoldt, Grabowski, Hickel,
  Neugebauer, Kresse, Janotti, and {Van de
  Walle}]{freysoldtFirstprinciplesCalculationsPoint2014}
Freysoldt,~C.; Grabowski,~B.; Hickel,~T.; Neugebauer,~J.; Kresse,~G.;
  Janotti,~A.; {Van de Walle},~C.~G. First-Principles Calculations for Point
  Defects in Solids. \emph{Reviews of Modern Physics} \textbf{2014}, \emph{86},
  253--305\relax
\mciteBstWouldAddEndPuncttrue
\mciteSetBstMidEndSepPunct{\mcitedefaultmidpunct}
{\mcitedefaultendpunct}{\mcitedefaultseppunct}\relax
\EndOfBibitem
\bibitem[Gaston \latin{et~al.}(2009)Gaston, Newman, Hansen, and
  {Lebrun-Grandi{\'e}}]{Gaston2009moose}
Gaston,~D.; Newman,~C.; Hansen,~G.; {Lebrun-Grandi{\'e}},~D. {{MOOSE}}: {{A}}
  Parallel Computational Framework for Coupled Systems of Nonlinear Equations.
  \emph{Nuclear Engineering and Design} \textbf{2009}, \emph{239},
  1768--1778\relax
\mciteBstWouldAddEndPuncttrue
\mciteSetBstMidEndSepPunct{\mcitedefaultmidpunct}
{\mcitedefaultendpunct}{\mcitedefaultseppunct}\relax
\EndOfBibitem
\bibitem[ahm()]{ahmadedl2024}
Ahmad, Zeeshan. 2024, edl\textunderscore solid GitHub repository,
  https://zenodo.org/doi/10.5281/zenodo.14538687\relax
\mciteBstWouldAddEndPuncttrue
\mciteSetBstMidEndSepPunct{\mcitedefaultmidpunct}
{\mcitedefaultendpunct}{\mcitedefaultseppunct}\relax
\EndOfBibitem
\bibitem[Pan \latin{et~al.}(2015)Pan, Cheng, and
  Qi]{panGeneralMethodPredict2015}
Pan,~J.; Cheng,~Y.-T.; Qi,~Y. General Method to Predict Voltage-Dependent Ionic
  Conduction in a Solid Electrolyte Coating on Electrodes. \emph{Physical
  Review B} \textbf{2015}, \emph{91}, 134116\relax
\mciteBstWouldAddEndPuncttrue
\mciteSetBstMidEndSepPunct{\mcitedefaultmidpunct}
{\mcitedefaultendpunct}{\mcitedefaultseppunct}\relax
\EndOfBibitem
\bibitem[Lin \latin{et~al.}(2020)Lin, Yong, Gustafson, Reedy, Ertekin,
  Krogstad, and Perry]{linDesignCationTransport2020a}
Lin,~Y.-Y.; Yong,~A. X.~B.; Gustafson,~W.~J.; Reedy,~C.~N.; Ertekin,~E.;
  Krogstad,~J.~A.; Perry,~N.~H. Toward Design of Cation Transport in
  Solid-State Battery Electrolytes: {{Structure-dynamics}} Relationships.
  \emph{Current Opinion in Solid State and Materials Science} \textbf{2020},
  \emph{24}, 100875\relax
\mciteBstWouldAddEndPuncttrue
\mciteSetBstMidEndSepPunct{\mcitedefaultmidpunct}
{\mcitedefaultendpunct}{\mcitedefaultseppunct}\relax
\EndOfBibitem
\bibitem[Ahmad \latin{et~al.}(2021)Ahmad, Venturi, Hafiz, and
  Viswanathan]{ahmadInterfacesSolidElectrolyte2021}
Ahmad,~Z.; Venturi,~V.; Hafiz,~H.; Viswanathan,~V. Interfaces in {{Solid
  Electrolyte Interphase}}: {{Implications}} for {{Lithium-Ion Batteries}}.
  \emph{The Journal of Physical Chemistry C} \textbf{2021}, \emph{125},
  11301--11309\relax
\mciteBstWouldAddEndPuncttrue
\mciteSetBstMidEndSepPunct{\mcitedefaultmidpunct}
{\mcitedefaultendpunct}{\mcitedefaultseppunct}\relax
\EndOfBibitem
\bibitem[Xiao \latin{et~al.}(2022)Xiao, Chen, and
  Maier]{xiaoDiscreteModelingIonic2022}
Xiao,~C.; Chen,~C.-C.; Maier,~J. Discrete Modeling of Ionic Space Charge Zones
  in Solids. \emph{Physical Chemistry Chemical Physics} \textbf{2022},
  \emph{24}, 11945--11957\relax
\mciteBstWouldAddEndPuncttrue
\mciteSetBstMidEndSepPunct{\mcitedefaultmidpunct}
{\mcitedefaultendpunct}{\mcitedefaultseppunct}\relax
\EndOfBibitem
\end{mcitethebibliography}

\includepdf[pages=1-3]{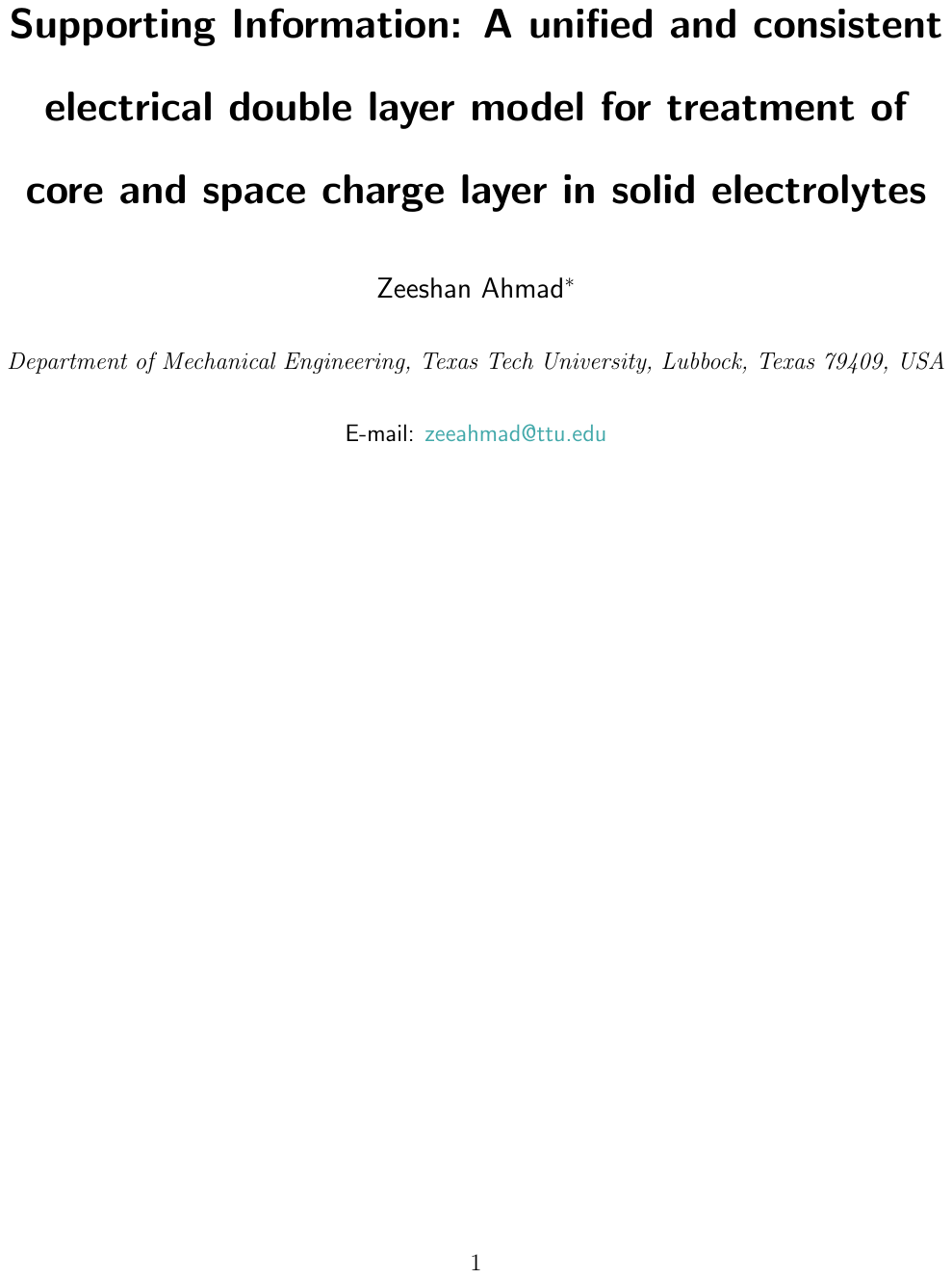}

\end{document}